
\def\pd{\partial}

\magnification=1200
\parskip 3 pt plus 1pt minus 1 pt
\rightline{DTP-92/11}
\rightline{February, 1992}
\vskip 2 true cm
\centerline{UNIVERSAL FIELD EQUATIONS WITH REPARAMETRISATION INVARIANCE}
\vskip 2.5 true cm
\centerline{D.B. FAIRLIE and J. GOVAERTS }
\vskip 0.5 true cm
\centerline{\it{Department of Mathematical Sciences}}
\centerline{\it{University of Durham, Durham DH1 3LE, England}}
\vskip 2 true cm
\centerline{Abstract}
 \vskip 1 true cm

New reparametrisation invariant field equations are constructed which
describe $d$-brane models in a space of $d+1$ dimensions. These equations,
like the recently discovered scalar field equations in $d+1$ dimensions,
are universal, in the sense that they can be derived from an infinity of
inequivalent Lagrangians, but are nonetheless Lorentz (Euclidean) invariant.
Moreover, they admit a hierarchical structure,
in which they can be derived by a sequence of iterations from an arbitrary
reparametrisation covariant Lagrangian, homogeneous of weight one.
None of the equations of motion which appear in the hierarchy of iterations
have
derivatives of the fields higher than the second.

The new sequence of Universal equations is related to the previous one  by
an inverse function  transformation. The particular case of $d=2$, giving a new
reparametrisation invariant string equation in 3 dimensions is solved.

\vfill\eject
\vskip 10pt
\centerline{\bf 1. Introduction}
\vskip 10pt
   In a recent paper with A. Morozov$^{[1]}$\rlap, we discovered an  equation
for a single field $\phi$ in $d$ dimensions with some remarkable properties.

\item{1)} It is invariant under field redefinitions.
\item{2)} It can be derived from an infinity of inequivalent Lagrangians
                                (hence is Universal).
\item{3)} It involves no derivatives higher than the second.
\item{4)} It is the penultimate step in an iterative hierarchy of equations,
          starting from any  Lagrangian, dependent only on first
          derivatives of the field  and homogeneous of weight one.

We have now found an equation for $d+1$ fields dependent on $d$ variables
which satisfy  2,3,4 above. The requirement 1 is replaced by reparametrisation
invariance in the $d$ base space variables.
Furthermore there is a transformation which takes
one equation into the other. The cases d=2,3 (a new string in 3 dimensions, a
new membrane in 4) have some physical importance, and may cast further light
upon problems of quantum gravity, and perhaps also classical gravity, as the
equations are related to the Plebanski equation$^{[2]}$.

The original Universal Field Equation in [1] takes the form
$$\det\pmatrix{0&\phi_1&\phi_2&\ldots&\phi_d\cr
               \phi_1&\phi_{11}&\phi_{12}&\ldots&\phi_{1d}\cr
               \phi_2&\phi_{12}&\phi_{22}&\ldots&\phi_{2d}\cr
                    .&\       .&\       .&\ddots&\       .\cr
               \phi_d&\phi_{1d}&\phi_{2d}&\ldots&\phi_{dd}\cr}=0,\eqno(1.1)$$
or, in short
$$ \det \pmatrix {0&\phi_{k}\cr\phi_l&\phi_{kl}\cr}.\eqno(1.2)          $$
Here $\phi_{k},\    \phi_{kl}$ denote partial derivatives
 ${\pd\phi\over\pd x_k },\ {\pd^2\phi\over\pd x_k\pd x_l }$ respectively.

 It is shown in [1] that equation (1.1) follows as the end result of the
following iterative construction of a hierarchy of equations, starting from an
arbitrary generic Lagrangian.

 Suppose ${\cal L}_0(\phi_j),\ j=1,\dots d$ is a Lagrangian which
depends only upon first derivatives
of the field $\phi$, and is  homogeneous of weight one, and the matrix
$M_{ij}=  {\pd^2{\cal L}_0(\phi_k)\over\pd\phi_i\pd\phi_j} $ is of maximal rank
$d-1$, which is the generic case.
Denote by ${\cal E}$ the Euler differential operator
$${\cal E}=-{\pd\over\pd\phi}
 +\pd_i {\pd\over\pd\phi_i}-\pd_i\pd_j{\pd\over\pd\phi_{ij}}\dots
\eqno(1.3)$$
(In principle the expansion continues indefinitely  but it is sufficient for
our purposes to terminate at the stage of second derivatives  $\phi_{ij}$).

Now consider the sequence of equations of motion;
$$\eqalign{ {\cal E\ L}_0 &=0\cr
            {\cal E\ L}_0{\cal E\ L}_0 &=0\cr
{\cal E\ L}_0{\cal E\ L}_0{\cal E\ L}_0 &=0\quad\hbox{etc.}\cr}\eqno(1.4)$$
Then this sequence terminates after $d$ iterations when the left hand side
vanishes identically. At the penultimate step the resulting equation of motion
is universal; i.e. is independent of the details of ${\cal L}_0$, and is in
fact
proportional to the equation (1.1).
 Among  the remarkable features of this
construction are the consequences that the equations involve nothing further
than second derivatives of the field $\phi$ and although the starting
Lagrangian need have no symmetry, the final equation is $GL(d)$ invariant.
If the homogeneity requirement is dropped, then the sequence results after
$d$ iterations with the equation, independently of the initial Lagrangian,
$$\det{\pd^2\phi\over\pd x_k\pd x_l }=0,\eqno(1.5)     $$
and terminates at the next stage in the iteration.

The very nature of our construction implies the existence of an infinite number
of conserved quantities for our Universal Equation which take the form of
inequivalent Lagrangians, (i.e. Lagrangians which do not differ by
divergences).
 Recently we succeeded in extending these results in the following manner;
we discovered a Universal Equation for $d+1$ fields in $d$ dimensions
with the property of covariance up to an overall factor under reparametrisation
of the base space. That there is only one such equation rather than $d+1$ is a
consequence of Ward identities for reparametrisation invariance.
\vskip 10pt
\centerline{\bf 2.  Reparametisation Invariance and Ward identities}
\vskip 10pt
Consider a Lagrangian density ${\cal L}(\phi^a_i,\phi^a_{ij}) $ for $\cal D$
fields dependent on $d$ co-ordinates such that
$${\cal L}(R_{ij}\phi^a_j,R_{ik}R_{jl}\phi^a_{kl}+T^k_{ij}\phi^a_k) =
(\det R_{ij})^\alpha
{\cal L}(\phi^a_i,\phi^a_{ij}) \eqno(2.1)$$
given any coefficients $R_{ij},\  T^k_{ij}=T^k_{ji}$.
In consequence of differentiation with respect to those parameters
we have the identities
$$
\eqalignno{\phi^a_j{\pd{\cal L}\over\pd\phi^a_i}(\phi^b_k,\phi^b_{kl})
+2\phi^a_{jk}{\pd{\cal L}\over\pd\phi^a_{ik}}(\phi^b_k,\phi^b_{kl})&=
\alpha\delta_{ij}{\cal L}(\phi^a_i,\phi^a_{ij})&(2.2a)\cr
 \phi^a_k{\pd{\cal L}\over\pd\phi^a_{ij}}(\phi^b_k,\phi^b_{kl})&=0.&(2.2b)\cr}
$$
Under a general reparametrisation of co-ordinates of the form
$$x_i\rightarrow y_i= y_i(x_j),\ \phi^a(x_i)\rightarrow \tilde\phi^a(y_i)=
\phi^a(x_i),\eqno(2.3)$$
we have
$$\eqalign { \tilde\phi^a_i=R_{ij}\phi^a_j,&\quad
\tilde\phi^a_{ij}  =R_{ik}R_{jl}\phi^a_{kl}+T^k_{ij}\phi^a_k\cr
\hbox{with}\quad R_{ij}={\pd  x_j\over\pd y_i},&\quad T_{ij}^k=
{\pd^2x_k\over\pd y_i\pd y_j}.\cr }   \eqno(2.4)    $$
The case $\alpha=1$ corresponds to a reparametrisation invariant action.

Given a Lagrangian density with the above properties, consider the action of
the Euler derivative;
$${\cal E}_a{\cal L}=\pd_i{\pd{\cal L}\over\pd\phi^a_{i}}-
\pd_i\pd_j{\pd{\cal L}\over\pd\phi^a_{ij}}.\eqno(2.5)$$
We then have
$$\eqalign{\phi^a_i{\cal E}_a{\cal L}=&
\phi^a_i\pd_j\{{\pd{\cal L}\over\pd\phi^a_{j}}-
 \pd_k{\pd{\cal L}\over\pd\phi^a_{jk}}\}\cr
=& \pd_j\{\phi^a_i{\pd{\cal L}\over\pd\phi^a_{j}}+
\phi^a_{ik}{\pd{\cal L}\over\pd\phi^a_{jk}}-
\pd_k\bigl[ \phi^a_i{\pd{\cal L}\over\pd\phi^a_{jk}}\bigr]\}\cr
 &-\phi^a_{ij}\{  {\pd{\cal L}\over\pd\phi^a_{j}}-
\pd_k{\pd{\cal L}\over\pd\phi^a_{jk}}\}\cr
=& \pd_j\{\phi^a_i{\pd{\cal L}\over\pd\phi^a_{j}}+
2\phi^a_{ik}{\pd{\cal L}\over\pd\phi^a_{jk}}\}-\pd_i{\cal L}-
\pd_j\pd_k\bigl[ \phi^a_i{\pd{\cal L}\over\pd\phi^a_{jk}}\bigr],\cr}
\eqno(2.6)$$
which, by using the identities (2.2a,\ 2.2b) reduces to
$$\phi^a_i{\cal E}_a{\cal L}=(\alpha-1)\pd_i{\cal L}.  \eqno(2.7)$$
 Whenever $\alpha =1$, corresponding to a reparametriation invariant
action, we have the associated $d$ Ward identities:
$$\phi^a_i{\cal E}_a{\cal L}=0. \eqno(2.8)$$
Consequently there are only ${\cal D}-d$ independent equations of motion.
\vskip 10pt
\centerline{\bf 3. The Case ${\cal D}= d+1$}
\vskip 10pt
    The situation ${\cal D} = d+1$ includes that of a string in 3 dimensions,
or a membrane in 4.
On account of reparametrisation invariance there is only one independent
equation of motion. It is convenient to define the Jacobians
$$ J_a =
(-1)^d\epsilon_{ab_1b_2\dots b_d}\phi^{b_1}_1\phi^{b_2}_2\dots \phi^{b_d}_d.
\eqno(3.1)$$
The following properties hold;
$$\eqalign{i)\quad& \sum_a   \phi^a_iJ_a=0\quad\forall i,\cr
ii)\quad&\sum_a\phi^a_i{\pd J_a\over \pd \phi^b_j}+\delta_{ij}J_b=0,\cr
 iii)\quad& {\pd J_a\over \pd \phi^b_i}=-{\pd J_b\over \pd \phi^a_i }.\cr}
\eqno(3.2)$$
The identities
$\phi^a_k{\pd{\cal L}\over\pd\phi^a_{ij}}(\phi^b_i,\phi^b_{ij})=0 $ show that
in
the present case we must have (no summation over $a$)
$${\pd{\cal L}\over\pd\phi^a_{ij}}(\phi^b_k,\phi^b_{kl})=J_aK^a_{ij}
(\phi^a_i,\phi^a_{ij}).\eqno(3.3)$$
This result suggests that reparametrisation invariant actions which depend upon
$\phi^a_{ij}$ must involve those variables in the combination
$\Phi_{ij}=\sum_a\phi^a_{ij}J_a$.
Indeed it turns out that the Universal field equation in this case is simply
$$ \det(\Phi_{ij})=0.\eqno(3.4)$$
\vskip 10pt
\centerline{\bf 4. A hierarchy for ${\cal D}= d+1$}
\vskip 10pt
Consider a Lagrangian density $ {\cal L}_0( J_a)$ such that
$$ {\cal L}_0(\lambda J_a) \ =  \lambda{\cal L}_0( J_a).\eqno(4.1)$$
The corresponding action is then reparametrisation invariant.
We also have;
$$\eqalignno{
 {\pd{\cal L}_0  \over\pd J_b}(J_a)J_b&={\cal L}_0( J_a),&(4.2a)\cr
 {\pd^2{\cal L}_0  \over\pd J_a\pd J_b}(J_c)J_b&= 0,&(4.2b)\cr
 {\pd^2{\cal L}_0  \over\pd J_a\pd J_b}(J_c)&=C_{ij}(\phi^c_k)\phi^a_i\phi^a_j,
 &(4.2c)\cr}$$
where $C_{ij}(\phi^c_k)=C_{ji}(\phi^c_k)$ and is given by
$$C_{ij}={1\over 2J_aJ_b}\big[{\pd^2{\cal L}_0\over\pd\phi^a_i\pd\phi^b_j}
+{\pd^2{\cal L}_0\over\pd\phi^a_j\pd\phi^b_i}\bigr],\quad\forall a,b,i,j.
\eqno(4.3)$$
The equations of motion are
$${\cal E}_a{\cal L}_0 = {\pd^2{\cal L}_0 \over\pd J_b\pd J_c}
{\pd J_b\over\pd\phi^a_i}{\pd
J_c\over\pd\phi^d_j}\phi_{ij}^d=J_aC_{ij}\Phi_{ij}
.\eqno(4.4)$$ This shows that the independent equation of motion is always
given
by
$${1\over J_a}{\cal E}_a{\cal L}_0 = C_{ij}\Phi_{ij},\quad\forall
a.\eqno(4.5)$$
This suggests that, in analogy to the case ${\cal D}=1$, $d$ arbitrary,
we consider the generating function
$\exp(\lambda{\cal L}_0{1\over J_a}{ \cal E}_a)\cdot{ \cal L}_0$
for any $a$ which
 generates term by term an iterative sequence of equations of motion.  This
sequence possesses the following remarkable properties:

\item{1)} It is a finite polynomial of order $d$ in $\lambda$.
\item{2)} It depends only upon $\phi^a_i,\  \phi^a_{ij}$, and no higher
          derivatives.
\item{3)} Every term in the polynomial is a Lagrangian
          density for a reparametrisation
          invariant action, which is multilinear in $\phi^a_{ij}$.
\item{4)} Up to a factor dependent on ${ \cal L}_0$, the coefficient of
          $\lambda^d$ is independent of the details of
          ${\cal L}_0$ and hence is  Universal. This is the equation (3.4).

This means that the Universal equation may be derived from an infinite number
of inequivalent Lagrangians, hence the terminology, and may therefore be
integrable. The proof of these statements will be merely sketched here. It
depends upon the fact that it turns out that this is not
 an independent
hierarchy of iterative equations but is a transformation of the original
case ${\cal D}=1$, $d$ arbitrary. Introduce a new variable  $y$ which is in the
nature of a co-ordinate, and consider ${\cal L}_0(J_a)$ where now
$\phi^a(x_i,y)$ depends upon the auxiliary co-ordinate, but $J_a$ is still
given by equation (3.1), i.e. it is the Jacobian of all the $\phi^b$ excluding
$\phi^a$ with respect to all the coordinates $x_i$, excluding $y$.
Then the equations of motion do not involve derivatives with respect to
$y$ as ${\cal L}_0$ depends only upon ${\pd \phi^a\over\pd x_i}$. Indeed
$$[\pd_{x_i}{\pd\over\pd\phi^a_i}+\pd_{y}{\pd\over\pd\phi^a_y}]{\cal L}_0
=\pd_{x_i}{\pd\over\pd\phi^a_i}{\cal L}_0(J_a)={\cal E}_a{\cal L}_0(J_a).
\eqno(4.6)
$$
Thus if ${\cal L}_0(J_a)$ is homogeneous of weight one, i.e. it satisfies
(4.1),
then the action is reparametrisation invariant in the $x$'s and takes the form
$${\cal S}=\int\prod_i dx_idy{\cal L}_0(J_a).\eqno(4.7)$$
Now perform the inverse function transformation; i.e. instead of treating
$\phi^a$  as dependent functions and $x_i,\ y$ as independent, think of
$x_i,\ y$ as described as functions of $\phi^a$.
We have then
$${\cal S}=\int\prod_idx_idy{\cal L}_0(J_a)\equiv
\int\prod_ad\phi^a{\cal L}_0({ \pd y\over\pd \phi^a}).\eqno(4.8)$$
Remarkably, on account of the homogeneity properties, all other jacobians
cancel in the above transformation of variables. But the transformed action is
just of the form of the starting action for the hierarchy discovered in [1].
A Lagrangian  ${\cal L}_0({ \pd y\over\pd \phi^a})$ of weight one dependent
upon only first derivatives has an equation of motion
   covariant under field redefinitions of $y$.
It can be taken as the beginning of an iterative sequence (1.4), which
culminates in the Universal field equation (1.1). Clearly it is possible to
map each member of the iterative sequence (1.4) into a new iterative sequence,
which this time, instead of invariance under field redefinitions, implements
reparametrisation invariance for each sequential action.
The remarkable feature of this procedure is that all derivatives with respect
to
$y$ disappear from the transformation of the generating function

$$\eqalign{
\bigl[\exp\lambda {\cal L}_0({\pd y\over\pd\phi^a}){\cal E}_y]{\cal L}_0
({\pd y\over\pd\phi^a})=& {\cal L}_0({\pd y\over\pd\phi^a})
\det\bigl[\delta_{ab}+\lambda{\pd^2{\cal L}_0\over\pd X_a\pd X_c}
{\pd^2{y}\over\pd \phi^c\pd \phi^b}\bigr],\cr
\hbox{where}\quad  X_a=&{\pd y\over\pd\phi^a}.\cr}\eqno(4.9)$$
Suppose this equation is re-expressed in terms of the inverse functions
$\phi^a(x_i,y)$. Then this results in the following transformed equations
$\forall a$;
$$\exp\bigl[\lambda {{\cal L}_0(J_a)\over J_a}{\cal E}_a\bigr]{\cal L}_0(J_a)
={\cal L}_0(J_a)\det\bigl[\delta_{ij}+\lambda C_{ik}(\phi^a_i)\Phi_{kj}\bigr]=
 {\cal L}(\phi^a_i,\phi^a_{ij},\lambda)
\eqno(4.10)$$
where ${\cal L}_0(\lambda J_a)=\lambda{\cal L}_0( J_a)$ and $C_{ik}(\phi^a_i)$
and $\Phi_{kj}$ are defined by equations  (4.3) and
$\quad\Phi_{ij}=\sum_a\phi^a_{ij}J_a$.
The remarkable feature of this transformation is that all possible dependence
of (4.10) on $\phi^a_y,\ \phi^a_{iy},\  \phi^a_{yy}$ has disappeared, so $y$ is
effectively a parameter as far as equation (4.10) is concerned. The equation
$${\cal E}_a {\cal L}(\phi^a_i,\phi^a_{ij},\lambda) ={J_a\over{\cal L}_0}
{\pd\over\pd\lambda} {\cal L}(\phi^a_i,\phi^a_{ij},\lambda)\eqno(4.11)$$
is the generating function for the reparametrisation invariant hierarchy which
terminates at the $d$th level, with the term in $\lambda^d$
being
$$
{\cal L}_0(J_a)\det[C_{ik}\Phi_{kj}]={\cal L}_0(J_a)\det[C_{ik}]\det[\Phi_{kj}]
.\eqno(4.12)$$
Since for generic
$ {\cal L}_0(J_a),\ \det[C_{ik}]\neq 0$, we have the Universal equation of
motion
$$\det[\Phi_{ij}]=\det[\phi^a_{ij}J_a]=0.\eqno(4.13)$$
\vskip 10pt
\centerline{\bf 5. Solutions for $d=2$, Duality}
\vskip 10 pt
This is the case of the alternative reparametrisation invariant
string in 3 dimensions.  It is apparently
different from that proposed by Grigore$^{[3]}$ since he does not consider
Lagrangians dependent on second derivatives. From the analysis presented in the
above section the solutions of the equation ${\rm det}[\Phi_{ij}]=0 $ are
equivalent
to solving the Universal field equation for one variable $y=f  $ in three
dimensions $X, Y,Z~^{[1]}$; viz:
$$      \det\pmatrix{0& f_X& f_Y&f_Z\cr
                     f_X&f_{XX}&f_{XY}&f_{XZ}\cr
                     f_Y&f_{XY}&f_{YY}&f_{YZ}\cr
                     f_Z&f_{XZ}&f_{YZ}&f_{ZZ}\cr}=0 .  \eqno(5.1)$$
Solutions of this equation correspond to {\it developable surfaces}$^{[4]}$.
Such surfaces, with the exception of the plane,
cone and cylinder are described by the {\it tangent surface of a curve}, i.e.
the locus of points on the tangents to an arbitrary curve. It is shown in [4]
that such surfaces are two sheeted, with a cusp at the curve in question.
Another parametrisation of such a surface is given by
$X=X(\sigma,\tau),\ Y=Y(\sigma,\tau),\ Z=Z(\sigma,\tau)$ , the string
formulation, which satisfies an equation of the form (4.13).
The cone and cylinder may be taken to represent closed string solutions, in
which a closed string propagates with and without expansion.
These two alternative
descriptions of the string, in a non parametric form as a surface, and the
parametrised form are reminiscent of the attempt of Morris$^{[5]}$
to develop  ordinary string theory in terms of intersections of surfaces.
  In this second form  the equation is related to Plebanski's equation for
self-dual Riemannian metrics$^{[2]}$.

Finally we should mention the second case in [1] which we found which admits a
formulation in terms of an infinite number of inequivalent Lagrangians.
It is the case of ${\cal D}= d$ fields in $d+1$ variables.
There are $d$ equations of
motion, invariant under field redefinitions of the form
$$\epsilon_{i_1 i_2 ... i_{d+1}}\epsilon_{j_1 j_2 ... j_{d+1}}
\prod^d_{r=1}\phi^{b_r}_{i_r}
\prod^d_{s=1}\phi^{c_s}_{j_s} \phi^a_{i_{d+1}j_{d+1}}.\eqno(5.2)$$
Under the transformation discussed previously,
these equations go over into the $d$ independent equations
$$ {\pd^2x_i\over\pd y^2}{\pd  x_j\over\pd y}=
{\pd^2x_j\over\pd y^2}{\pd  x_i\over\pd y},\quad\forall  i,j\eqno(5.3)$$
These equations are covariant under reparametrisations in the variable $y$.
They may be regarded as a dual version of the ${\cal D}=d+1,\  d=d$ theory
presented here.
\vskip 10pt
\leftline{\bf 6. Conclusions.}
\vskip 10pt
This paper and its predecessor$^{[1]}$ have revealed the existence of field
theory models which possess many interlinked  and unexpected properties;
reparametrisation invariance, infinitely many conservation laws,
absence of higher derivatives, and the hierarchy property. It is not at all
clear at this stage which is fundamental, and which are inevitable
consequences. Certainly the fact that these equations may be derived from
an infinity of Lagrangians, inequivalent in so far as they are not related by
divergences makes them interesting objects of study in their relation to
topological field theories$^{[6]}$ and variational procedures$^{[7]}$.
It is also not clear at the moment whether there is any connection with the
geometrical realisations of $W$ algebras$^{[8]}$, or indeed gravity
theories$^{[9]}$.

The extension of those ideas to arbitrary choices of target space and base
space  dimension remains elusive. It is relatively easy to write equations with
the correct covariance or reparametrisation properties, but there is a
superfluity of such equations (3 instead of 2 for ${\cal D}=4,\ d=2$ ),
and there is no  sign of the hierarchy property.
Nonetheless the solution space is non empty. The success of the
${\cal D}=3,\ d=2$ model may owe its existence to the  invariant
$$\epsilon_{abc}\epsilon_{ik}\phi^a_i\phi^b_j\phi^c_{kj}.\eqno(6.1)
$$
We hope to return to those questions and give a more detailed account of the
results of the present paper in the near future.
\vskip 10pt
\leftline{\bf Acknowledgement}
\vskip 5pt
This research was supported by the S.E.R.C. award of a Research
Assistantship to J.Govaerts.
\vfill\eject
\centerline{\bf REFERENCES}
\frenchspacing
\item{[1]}Fairlie D. B., Govaerts J. and Morozov A., Universal Field Equations
with Covariant Solutions, Durham preprint DTP-91/55, hepth-9110022 (October
1991), to appear in {\it Nuclear Physics B}.
\item{[2]}Finley J. D. and Plebanski J. F., {\it Jour. Math. Phys.}\ {\bf 17}\
(1976)\ 585.
\item{[3]}Grigore D. R., A Derivation of the Nambu-Goto Action from
Invariance Principles, preprint CERN-TH.6101/91 (May 1991).
\item{[4]}Eisenhart L. P., {\it An Introduction to Differential Geometry},
Princeton University Press (1940) 59.
\item{[5]}Morris T. W., {\it Phys. Lett.}\ {\bf B 202}\ (1988)\ 222;
 \ {\it Nucl. Phys.}\ {\bf B 331}\ (1990)\ 694.
\item{[6]}Witten E., {\it Comm. Math. Phys.}\ {\bf  117}\ (1988)\ 353;
ibid.\ {\bf 118}\ (1988)\ 411.
\item{[7]}Olver E., {\it Applications of Lie Groups to Differential Equations}
\ Graduate Texts in Mathematics, {\bf 107}, Springer Verlag (1986), p. 252.
\item{[8]}Gervais J.-L. and Matsuo Y., preprint LPTENS-91/35, NBI-HE-91/50,
hepth-9201026 (January 1992).
\item{[9]}Hull C. M., The Geometry of $W$-Gravity, preprint QMW/PH/91/6 (June
1991).
\end